\documentclass[journal, a4paper]{IEEEtran}
\IEEEoverridecommandlockouts
\ifCLASSINFOpdf
\usepackage[pdftex]{graphicx}
\else
\fi
\usepackage{amsmath}
\usepackage{array}
\ifCLASSOPTIONcompsoc
\usepackage[caption=false,font=normalsize,labelfont=sf,textfont=sf]{subfig}
\else
\usepackage[caption=false,font=footnotesize]{subfig}
\fi
\usepackage{fixltx2e}
\usepackage{stfloats}
\usepackage{cite}
\usepackage{url}
\usepackage{hyperref} 
\hypersetup{colorlinks = true, linkcolor=magenta,
	citecolor=red, urlcolor=blue} 

\usepackage{amssymb}
\usepackage{amsthm}
\usepackage{algpseudocode}
\usepackage{algorithm}

\usepackage{graphicx}
\usepackage{mathptmx}
\usepackage{amsmath}
\usepackage{xcolor}
\allowdisplaybreaks

\newtheorem{prop}{Proposition}

\theoremstyle{definition}

\newcommand{\beq}{\begin{equation}}
\newcommand{\eeq}{\end{equation}}
\newcommand{\beqn}{\begin{eqnarray}}
\newcommand{\eeqn}{\end{eqnarray}}
\newcommand{\beqno}{\begin{eqnarray*}}
	\newcommand{\eeqno}{\end{eqnarray*}}
\newcommand{\bma}{\begin{displaymath}}
\newcommand{\ema}{\end{displaymath}}
\newcommand{\bnu}{\begin{enumerate}}
	\newcommand{\enu}{\end{enumerate}}
\newcommand{\bce}{\begin{center}}
	\newcommand{\ece}{\end{center}}
\newcommand{\btb}{\begin{tabular}}
	\newcommand{\etb}{\end{tabular}}

\usepackage{setspace}

\begin{document}
\title{User Preference Learning Aided Collaborative Edge Caching for Small Cell Networks}

\author{\IEEEauthorblockN{Md Ferdous Pervej\IEEEauthorrefmark{1}, Le Thanh Tan\IEEEauthorrefmark{2}, and Rose Qingyang Hu\IEEEauthorrefmark{3}}\\
\IEEEauthorblockA{\IEEEauthorrefmark{1}Department of Electrical and Computer Engineering, North Carolina State University, Raleigh, NC 27695, USA} \\
\IEEEauthorblockA{\IEEEauthorrefmark{2}Commonwealth Cyber Initiative, Old Dominion University, Norfolk, VA 23529, USA} \\
\IEEEauthorblockA{\IEEEauthorrefmark{3}Department of Electrical and Computer Engineering, Utah State University, Logan, UT 84322, USA} \\
Email: \tt mpervej@ncsu.edu, tle@odu.edu, rose.hu@usu.edu 

\thanks{The work of M. F. Pervej, L. T. Tan and R. Q. Hu were supported in part by National Science Foundation under grants NeTS 1423348 and EARS 1547312 as well as in part by the Intel Corporation.}
\thanks{This is the \textbf{technical report} of \cite{Pervej_Cost}.}}

\maketitle

\begin{abstract}
While next-generation wireless networks intend leveraging edge caching for enhanced spectral efficiency, quality of service, end-to-end latency, content sharing cost, etc., several aspects of it are yet to be addressed to make it a reality. 
One of the fundamental mysteries in a cache-enabled network is predicting what content to cache and where to cache so that high caching content availability is accomplished. 
For simplicity, most of the legacy systems utilize a static estimation - based on Zipf distribution, which, in reality, may not be adequate to capture the dynamic behaviors of the contents popularities. 
Forecasting user's preferences can proactively allocate caching resources and cache the needed contents, which is especially important in a dynamic environment with real-time service needs. 
Motivated by this, we propose a long short-term memory (LSTM) based sequential model that is capable of capturing the temporal dynamics of the users' preferences for the available contents in the content library.
Besides, for a more efficient edge caching solution, different nodes in proximity can collaborate to help each other. 
Based on the forecast, a non-convex optimization problem is formulated to minimize content sharing costs among these nodes. 
Moreover, a greedy algorithm is used to achieve a sub-optimal solution. 
Using extensive simulation and analysis, we validate that the proposed algorithm performs better than other existing schemes.
\end{abstract}

 \vspace{-0.1 in}

\IEEEpeerreviewmaketitle

\section{Introduction}
Wireless user penetration is consistently increasing with a continuous emergence of new and sophisticated user-defined applications.
This steers wireless technologies to evolve rapidly from one generation to the next generation striving to soothe the yearning for more enhanced spectral efficiency, energy efficiency, quality of experience, operation cost, etc. 
Even though the existing wireless networks ensured a very promising performance in these contexts, new demands on capacity and other performance have never ceased to emerge \cite{5978416}.
Besides, with the advent of the Internet of everything \cite{bandyopadhyay2011internet, nayak20206g}, the incompetence of these legacy technologies became more apparent.
Therefore, researchers are continuously in a toiled search for new technologies that can be adopted on top of the existing ones for future generation networks. 
Among many other impressive ideas, moving towards the user-centric distributed network infrastructure is a promising one \cite{8447267, 8618350, 8410473, 8169053, ye2018smart, pervej2020eco, pervej2020dynamic}.

Note that a user-centric network platform significantly reduces energy consumption \cite{pervej2020eco}, increases network throughput \cite{pervej2020dynamic} as well as enhances the utilization of the much-needed spectrum \cite{8447267, 8618350, 8410473}.
On the other hand, edge caching is the concept of storing popular contents close to the end users.
Therefore, leveraging edge caching, user-centric network infrastructure efficiently utilizes the network bandwidth and significantly reduces the congestion on the links to the centralized cloud server \cite{8447267, 8618350, 8410473}.
Besides, edge caching is considered as a promising scheme to support video applications due to their traffic volume escalation and stringent QoS requirements \cite{6787081}.
To mitigate this bottleneck, one of the prominent advocacy of caching is to alleviate the bandwidth demand in the centralized network segments by storing the popular video contents in the local/nearby nodes.

In the literature \cite{8618350, 8410473, 8169053, 6600983,7944647, 8977517}, several researchers studied edge caching in terms of different performance metrics.
Tan \textit{et al.} conducted static popularity based throughput maximization analysis in \cite{8410473}.
A novel content delivery delay minimization problem was studied in \cite{8169053}.
Shanmugam \textit{et al.} \cite{6600983} also considered both coded and uncoded cases for caching contents at the helper nodes to minimize the content downloading time. 
Song \textit{et al.} \cite{7944647} proposed a dynamic approach for the scenarios of the single player and the multiple players. 
Recently, Jiang \textit{et al.} considered a cache placement strategy to minimize network costs in \cite{8977517}.

In this work, we develop a new caching platform that allows user caching, device to device (D2D) communications and collaborations among edge caching nodes. 
Furthermore, long short-term memory (LSTM) based sequential model is proposed for the content popularity estimation, where the dynamic nature of the content's popularity would be perceived.
The contributions in this paper are summarized as follows.

\begin{enumerate}
	\item To capture the short-temporal user dynamics, we use LSTM for forecasting user preferences ahead of time.
	\item To fully exploit the advantages of edge caching, we propose a collaborative communication framework, in which different nodes in the same cluster can share contents among each other.
	\item We formulate the optimization problems to minimize the content sharing costs under the constraints of limited and dynamic storage capacities at both the users and the BSs for both heterogeneous caching placement and homogeneous caching placement scenarios.
	We then analyze the content sharing cost and develop collaborative edge caching algorithms to configure the parameters of caching placement.
	\item Numerical results are illustrated to validate the theoretical findings and the performance gain of our proposed algorithms.
\end{enumerate}

The outline of this paper is as follows. Section~\ref{SYS_MOD} describes the system model. Section~\ref{Dynamic_User_Preference_Prediction} presents our dynamic user preference prediction. In Section~\ref{Caching_Model}, we introduce our caching model and optimization problems. We perform analysis and present our algorithms in Section~\ref{Observations_Joint_Solver}. Section~\ref{Result} presents the performance results followed by the concluding remarks in Section~\ref{Conclu}.

\section{System Model}
\label{SYS_MOD}
This section introduces our proposed user-centric system model, followed by the dynamic user preference prediction model.

\begin{figure}
	\centering
	\includegraphics[width= 0.5 \textwidth]{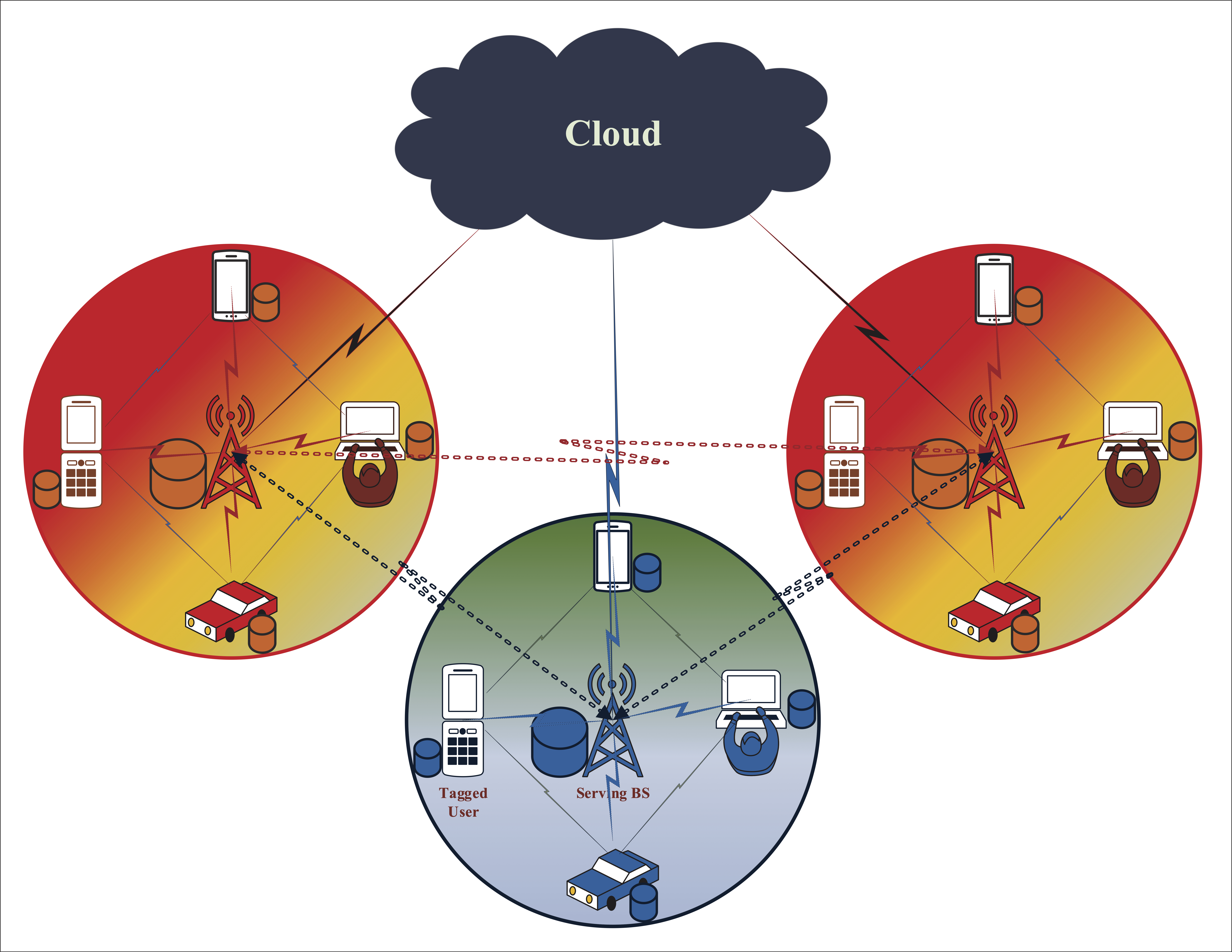}
	\caption{System Model for Collaborative Caching}
	\label{sysmod}
\end{figure}

\subsection{User-Centric System Model}
A set of D2D users $\mathcal{U} = \{i\}$, $i \in \{1,2,\dots,U\}$, are distributed in the coverage area of the BSs; and $C_d$ is the caching size of all the  users.
Considering a cluster based system model, we assume each cluster consists of a number of BSs\footnote{In each cluster, different BSs use orthogonal bandwidth so there is no interference within a cluster}.  In each cluster, there are an equal number of BSs, each of which has an equal caching capacity. 
Here, $\mathcal{B} = \{j\}$, where $j \in \{1,2,\dots,B\}$ and $C_b$ represent the set of BSs and the cache storage size, respectively.
For simplicity, we assume that each BS  serves an equal number of users. 
We denote a D2D requesting node and the serving BS as the tagged D2D node and the tagged BS\footnote{Throughout this paper, the name serving BS and tagged BS are used interchangeably.}, respectively.
Furthermore, all the D2D nodes in a single cell are assumed to be in the communication range of each other.  
All the D2D users are in the communication range with at least one of the  BSs. 
We take $F$ most popular contents, in our content catalog, where $\mathcal{F} = \{k\}$ and $k \in \{1,2,\dots,F\}$. 
Note that this assumption of fixed content is made only during a period.
Considering the age of information (AoI) and content freshness, similar to \cite{8447267, 8618350, 8647483}, we assume that new popular content is added periodically removing the least popular ones.
Furthermore, following the widely used notion, we assume that all the contents have the same size, which is denoted by $S_f$. 
In the case that the content sizes are different, we can divide the content into equal segments of packets and then store those segments \cite{7775114,7524381}.

The proposed model must be carefully designed to satisfy the critical latency of real-time communication by delivering the requested contents from the local cache as much as possible. If a tagged user needs to access the desired content, before sending the content request to other nodes, it firstly checks its own cache storage. 
The tagged user sends the content request to the neighboring D2D nodes that are residing in the same cell and are within the communication range, if the requested content is not found in its own cache storage. 
If the content is available in one of the neighboring D2D nodes, the content can be directly served from that node to the tagged user. 
If none of the D2D nodes has the requested content, the request is then forwarded to the serving BS, which delivers content to the tagged user if the content is found its storage. 
If the requested content does not exist in the serving BS's cache, the serving BS forwards the request to the neighboring BSs residing in the same cluster. 
If the content is not available in any of these stores, it can be downloaded from the cloud, which is considered as expensive consuming of time and bandwidth.

In this paper, our problem formulation is modeled in two steps. 
The first step models the dynamic content preferences of the users. 
The novel intention of this paper is to model the per-user content preference in a dynamic manner. 
The second step performs the caching policy based on the prediction. 
The  goal is to store the most probable `to be requested' contents in the future time slots. 
Using the actual requests of the users, we present the optimization model aiming for minimizing the total cost  of content sharing. 
In the next section,  prediction model is presented.

\section{Dynamic User Preference Prediction}	
\label{Dynamic_User_Preference_Prediction}

In this section, we firstly define the terms that are used throughout this paper to avoid any confusions of cross-domain nomenclatures. 
The proposed approach for modeling the dynamic user preference is also presented.

\subsection{Definitions}
Modeling content popularity and heterogeneous preferences in a system model as presented in Fig. \ref{sysmod} is always challenging. 
If we only take content popularity into account, it dynamically varies over time and locations, let alone user preferences. Motivated by this, we introduce different terms to facilitate the  understanding of  different aspects. 

\subsubsection{Content Popularity} 
In a general sense, content popularity defines the fondness of a content. 
More formally, the content popularity is the probability distribution of the content, which expresses the number of times a content $k \in \{\mathcal{F} \}$ is accessed or requested by all the users. 
If we consider only a small geographic region such as a small cell, this is usually noted as local popularity.
In most of the legacy networks, Zipf distribution has been widely used to model the content popularity \cite{6600983,cha2007tube}. 
The probability mass function (pmf) of this Zipf distribution is represented by 
\begin{equation}
\begin{aligned}
	\mathrm{p}_{f_k} = \frac{k^{-\gamma}}{\sum_{k=1}^{F} k^{-\gamma}},
\end{aligned}
\label{Zipf_pmf_upledge}
\end{equation}
where $F$ denotes the number of content, while $\gamma$ denotes the skewness of the content popularity.  

\subsubsection{User Content Preference}
Rather than considering the fondness of a content to all the user, when we consider per user basis, we denote the term as the user content preference. 
Formally, the user content preference  defines the probability of requesting a content $f_k$ by a user $u_i$ given that the user actually makes a request. It is mathematically expressed as 
\begin{equation} 
	\mathbf{q}_{f|u_i}(t) = \left[q_{f_1|u_i}(t), q_{f_2|u_i}(t), \dots , q_{f_F|u_i}(t)\right]^T, ~~ \forall i \in \{\mathcal{U}\},
\end{equation}
where $q_{f_k|u_i}(t)$ represents the probability that user $u_i$ requests content $f_k$ at time slot $t$ given that it actually makes a request. 
It is readily understandable that at a time slot $t$, we have $\sum_{k=1}^{F} q_{f_k|u_i}(t) = 1$. 
Furthermore, this is equivalent to $q_{f_k|u_i}(t) \stackrel{\Delta}{=} P_t(f_k|r_i)$, where $P_t(r_i(t))$ is the probability of event that user $u_i$ makes a request at time slot $t$. 
Note that in the time slot $t$, we may stack the user preference in a matrix for the ease of convenience as follows:
\begin{equation}
\begin{aligned}
\label{Content_Pref_Matrix}
	\mathbf{Q}_{u|f}^{U \times F}(t)	&= \begin{bmatrix}
	\mathbf{q}_{f|u_1}(t) & \mathbf{q}_{f|u_2}(t) & \dots & \mathbf{q}_{f|u_U}(t) 
	\end{bmatrix}^T.
\end{aligned}		
\end{equation}

\subsubsection{Activity Level of User}

As the name suggests, we define the probability that a user sends a content request as its activity level. 
We denote this by $r_i(t) \stackrel{\Delta}{=} P_t(r_i(t))$, where $\sum_{i = 1}^{U} P_t(r_i(t)) = 1, \forall u_i \in \{\mathcal{U}\}$. 
Moreover, we may stack all the elements for all users into a vector, which is $\mathbf{r}(t) = [r_1(t), r_2(t), \dots, r_U(t)]^T$.
Before expressing the mathematical equation of $r_i(t)$, let us define the user-content matrix for time slot $t$, which is given as
\begin{equation}
	\begin{aligned}
	\label{User_Content_Matrix}
	\mathbf{N}_{uf}^{U \times F}(t)	&= \begin{bmatrix}
	\mathbf{n}_{u_1,f}(t) & \mathbf{n}_{u_2,f}(t) & \dots & \mathbf{n}_{u_U,f}(t) 
	\end{bmatrix}^T, 
	\end{aligned}		
\end{equation}
where $\mathbf{n}_{u_i,f}(t) = [n_{u_i,f_1}(t) , n_{u_i,f_2}(t) , \dots , n_{u_i,f_F}(t)]^T$. Here, $n_{u_i,f_k}(t)$ denotes the number of incidents in which user $u_i$ has requested content $f_k$ in time slot $t$. Moreover, the following is written accordingly to represent the total number of requests made by all users for all contents in that time slot. 
\begin{equation}
\label{total_request}
q(t) = \sum_{i=1}^{U} \sum_{k=1}^{F} n_{u_i,f_k}(t).
\end{equation}

From the user-content matrix in (\ref{User_Content_Matrix}), we also define the following two terms. 
Let $n_{u_i}(t)$ be the summation of the content requests made by a user $u_i$ for all the contents $f_k \in \{\mathcal{F}\}$ in time slot $t$. 
Also, let $n_{f_k}(t)$ be the total number of requests for a particular content $f_k$ in that time slot by all users $u_i\in \{\mathcal{U}\}$. 
Then, these terms can be written as
\beqn
	n_{u_i}(t) & = \sum_{k = 1}^{F} n_{u_i,f_k}(t), ~~ \forall i \in \mathcal{U}, \label{No_of_Content_Reuquested_by_user}\\
	n_{f_k}(t) & = \sum_{i = 1}^{U} n_{u_i,f_k}(t), ~~ \forall k \in \mathcal{F}. \label{No_of_request_for_content_by_all_user}
\eeqn

We now express the activity level of the user as
\begin{equation}
\label{Activity_Level_of_User}
r_i(t) = \frac{n_{u_i}(t)}{q(t)}. 
\end{equation}
Recall that the motivation for introducing activity level is essential for knowing the load coming from each individual user. Furthermore, the conditional probability $q_{f_k|u_i}(t)$ that a user requests content $f_k$ given that it actually makes a request is written as
\begin{equation}
\label{Conditional_Prob_making_req}
q_{f_k|u_i}(t) = \frac{n_{u_i, f_k} (t)}{n_{u_i}(t)}. 
\end{equation}

Based on (\ref{Activity_Level_of_User}) and (\ref{Conditional_Prob_making_req}), the joint probability of the event that user $u_i$ actually makes the request and the requested content is $f_k$ at time slot $t$ is calculated as
\begin{equation}
\label{User_Content_Joint_Probability}
\begin{aligned}
q_{u_i,f_k} (t) &= r_i(t) q_{f_k|u_i}(t) \stackrel{\Delta}{=} P_t(r_i(t)) q_{f_k|u_i}(t).
\end{aligned}
\end{equation}
According to (\ref{User_Content_Joint_Probability}), it is readily observed that we need to predict the activity level of users as well as to predict which content the users request. 
Furthermore, we assume that we know the complete data, using the definitions, we have
$\sum_{i=1}^{U} r_i(t) q_{f_k|u_i} (t) = p_{f_k} (t)$, 
where $p_{f_k} (t)$ represents the global content popularity confined on that region for time slot $t$.

\subsection{Predicting Dynamic User Preferences Using LSTM}
In this subsection, we present a special kind of recurrent neural network (RNN), namely the LSTM, which is developed to avoid the long term dependencies in the RNN.
Usually, the structure of LSTM includes three gates, namely forget gate, input gate and output gate.
To model dynamic user preferences, a historical dataset is required. 
However there is no real dataset for modeling individual user behavior \cite{8425746}. 
Hence, at first, a synthetic dataset is generated for modeling dynamic user behavior. 
Note that, content popularity is usually modeled by using Zipf distribution, in which the parameter of skewness controls the distribution. 
Modeling the content popularity at each individual user level is different from modeling global content popularity. Therefore, a random skewness is used for all the users while generating the dataset. 
Firstly, a random content index order is generated for each user. 
Then, the number of requests for the user is generated by using random skewness in the allowable range of skewness. 
The governing equation for this is given below. 
\begin{equation}
\label{Random_Skewness}
\begin{aligned}
\gamma &= \left(\gamma^{max} - \gamma^{min}\right) \times \mathrm{Uniform}(0,1) + \gamma^{min}, 
\end{aligned}
\end{equation}
where $\gamma^{max}$ and $\gamma^{min}$ are the maximum and minimum values of skewness.
Note that, different content order and skewness are considered for each user. 
While the content's order is taken randomly for each user, the skewness is controlled by (\ref{Random_Skewness}). 
Furthermore, this value is only the initial value. correlated data are generated afterwards. 
The detailed procedure of generating the synthetic dataset is given in Alg.~ \ref{Synthetic_Data_Set_Generation}.

\begin{algorithm} 
\small
	\caption{Generating Synthetic Data Set}	
	\begin{algorithmic}	[1]
		\State select total number of users $U$, total number of content $F$ and total number of historical time slots, $t$
		\For {each user, $u_i \in \{U\}$}
		\State select random content index order, $\forall~f_k\in\{\mathcal{F}\}$  \label{Random_Order}
		\State $\mathrm{\gamma} \leftarrow \mathrm{Uniform}(\gamma_{\mathrm{min}}, \gamma_{\mathrm{max}})$ \Comment{use equation (\ref{Random_Skewness})}
		\State $\mathcal{N}_{\mathrm{int}}^{\mathrm{req}} \leftarrow \mathrm{Uniform}(\mathcal{N}_{\mathrm{min}}^{\mathrm{req}}, \mathcal{N}_{\mathrm{max}}^{\mathrm{req}})$ 
		\State $\mathrm{P}_{int} \leftarrow \mathrm{Uniform}\left((0,1), \mathcal{N}_{\mathrm{int}}^{\mathrm{req}}  \right)$  \label{Uniform_N_req} \Comment{$\mathcal{N}_{\mathrm{int}}^{\mathrm{req}}$ Uniform random numbers $\in \{0,1\}$}
		\State calculate $P_{f_k}(k=i)$ using equation (\ref{Zipf_pmf_upledge}) \label{Zipf_pmf_Alg}
		\State $P_{f_k} \leftarrow \sum_{j=1}^{k} P_{f_j}$ \Comment{cumulative sum of (\ref{Zipf_pmf_upledge})}  \label{cumsum}
		\State generate $\mathcal{N}^{\mathrm{req}}$ using $\mathrm{Histcounts}$$\left(\mathrm{P}_{int}, P_{f_k} \right)$ \Comment{$P_{f_k}$ of step (\ref{cumsum}), $\mathcal{N}^{\mathrm{req}} \in \mathbb{R}^{F}$}
		\For {$\forall$ $f_k \in \{\mathcal{F}\}$}
		\State sort $\mathcal{N}^{\mathrm{req}}$ as per initial generated index in step (\ref{Random_Order}) \Comment{generation in steps (\ref{Zipf_pmf_Alg} - \ref{cumsum}) is not sorted}
		\EndFor
		\EndFor 
		\State\textbf{return}: $\mathbf{N}_{uf}^{U \times F}(t) = \mathcal{N}^{\mathrm{req}}$, \label{int_generated_no}  \Comment{initial user content matrix, $t=1$}
		\For {each $t > 1$}
		\For {all user}
		\State $\mathcal{N}_{\mathrm{new}}^{\mathrm{req}} (t)$ using the initial generated number in step (\ref{int_generated_no})
		\If {$\mathcal{N}_{\mathrm{new}}^{\mathrm{req}} (t)$ $\neq$ $INT$}  \Comment{$INT$ refers to integer}
		\State $\mathcal{N}_{\mathrm{new}}^{\mathrm{req}} (t) \leftarrow$ $\mathrm{round}$$\left(\mathcal{N}_{\mathrm{new}}^{\mathrm{req}} (t) \right)$
		\EndIf
		\EndFor
		\State \textbf{return}: synthetic user-content data matrix $\mathbf{N}_{uf}^{U \times F}(t)$ 
		\EndFor	
	\end{algorithmic} \label{Synthetic_Data_Set_Generation}
\end{algorithm}

Given the historical dataset for $t \in \{1,2,\dots, N\}$ time slots, the next focus is on the LSTM based prediction model. 
A time ahead forecast of the probability of making a content request, i.e. activity level, and what content a user will request at the $t \in \{N+1\}$s time slots are also conducted. 
The model needs to be trained using the available historical data of the first $N$ time slots. 
Then, $(N+1)^{th}$ time slot's data needs to be predicted. 
The detailed procedure is discussed in what follows.

For the training, at  $t$, an entire row is fed to the input of the LSTM block meaning that the input $X_t$ of the LSTM is an entire row of the user-content matrix obtained from Alg. ~\ref{Synthetic_Data_Set_Generation}.
In (\ref{User_Content_Matrix}), the data of an individual user is fed to the LSTM model. 
Therefore, this has to be performed for all the users $u_i$, $i \in \mathcal{U}$. 
After the model is trained, the value of each row of the user-content matrix, $\mathbf{N}_{uf}^{U \times F}$ is forcasted for $t=(N+1)^{th}$ time slot. These forcasted value at  $t = N+1$ is next used to forecast the next time slot's data. 
Note that as the generated number is the prediction of how many times a user will request a content, which is not non-positive or fractional. 
However, the forecast results following the LSTM model may contain fractional value, which would be avoided by using $\mathrm{rounding}$. 
The detailed procedure is in Alg.~\ref{Predicting_values_Using_LSTM}.

\begin{algorithm} [!t]
\small
\caption{Predicting Sequential Data Using LSTM}
\begin{algorithmic}[1]
	\For {each cell, $j \in \mathcal{B}$}
		\For {each user, $u_i \in \{U_{b_j}\} $}
			\State take generated historical dataset from Alg. ~\ref{Synthetic_Data_Set_Generation}
			\State process the data to take the entire row for the time slot as input elements of the LSTM input
			\State divide the dataset into training, validation and test part 
			\State feed the data to the LSTM model \label{LSTM_first_blck}
			\State using the model forecast the value of the entire row for $(N+1)^{th}$ time slot
			\State save the trained model and store the values
		\EndFor
	\EndFor
	\State {\textbf{return}: predicted value for $(N+1)^{th}$ time slot}
\end{algorithmic} \label{Predicting_values_Using_LSTM}
\end{algorithm}

After running Alg.~\ref{Predicting_values_Using_LSTM}, $\hat{r}_i(N+1)$ and $\hat{q}_{f_k|u_i}(N+1)$ are calculated. Remember that $r_i(t)$ denotes the probability that a user actually makes the request, while $q_{f_k|u_i}(t)$ denotes the conditional probability that user $u_i$ request for content $f_k$ conditioned on $r_i(t)$. 
The predicted activity levels of the users are calculated as follows:
\begin{equation}
\label{Predicted_Activity_Level}
\hat{r}_i(t) = \frac{\hat{n}_{u_i}(t) }{\hat{q}(t)},
\end{equation}
where $\hat{n}_{u_i}(t) = \sum_{k = 1}^F \hat{n}_{u_i,f_k} (t)$, $\hat{q}(t) = \sum_{i = 1}^U \sum_{k = 1}^F \hat{n}_{u_i,f_k} (t)$ and $t = N+1$.

The predicted conditional probability that a user request for content $f_k$ given that she actually makes a request is calculated as
\begin{equation}
\label{Predicted_Conditional_Probability}
\hat{q}_{f_k|u_i}(t) = \frac{\hat{n}_{u_i,f_k}(t) }{\hat{n}_{u_i}(t)}.
\end{equation}
Thus, the predicted joint probability that a user will make a request for content $f_k$ is calculated as
\begin{equation}
\label{Predicted_Joint_Probability}
\hat{q}_{f_k,u_i}(t) = \hat{r}_i (t) \hat{q}_{f_k|u_i} (t).
\end{equation}
Without loss of generality, the preference probability of a user for the next time slots, $t=(N+1)$s, are calculated from this joint probability. 
Since $\hat{q}_{f_k, u_i}(t)$ is the joint probability, necessary normalization may require to keep the summation of these probabilities to be equal to $1$.

Note that the preference probability, $\hat{q}_{f_k,u_i}(t)$, can be modeled for all the future time slots, $\forall ~t \geq N+1$. This forecast period is completely on the system administrator's hand. 
Based on the requirements, it can be set to any reasonable time window. 
Furthermore, if the per time scale analysis is required, it can be easily modeled by considering the per time slot user's preference probabilities. 
However, in this works, the long term content caching probabilities are analyzed. 
As placing the content for each forecast time window may not be cost-efficient, the long term request probability is considered.  Thus, without loss of any generality, assuming the forecast window as fixed, the future content preferences of the users are considered as the average of the predicted $\hat{q}_{f_k,u_i}(t)$, $\forall ~t \geq N+1$. 

Let $N^{\mathrm{opt}}$ denote this fixed time window chosen by the network administrator. 
Then, the average $\hat{q}_{f_k,u_i}(t_o)$, $\forall~t_o \in \{N+1, N+2, \dots, N+N^{\mathrm{opt}}\}$ is considered as the preference probability of the user $u_i$ for evaluating the system performance\footnote{$t_o$ represent only the optimization time slots, while $t$ represent all time slot.}. 
Let $\rho_{f_k}^{u_i}$ denote this preference probability that has to be used for the performance evaluation. 
This then can be calculated as 
\begin{equation}
\label{Predicted_Preference}
	\rho_{f_k}^{u_i} = \frac{\sum_{t_0 = N+1}^{N+N^{\mathrm{opt}}}\hat{q}_{f_k,u_i} (t_0)}{N^{\mathrm{opt}}} ~~, \forall~ i \in \mathcal{U}~~ \text{and} ~~k \in \mathcal{F}.
\end{equation}
Since the predicted values of what content a user will request in the next time slot and what load it will create for the network are already known at this point, we now focus on the caching placement.

\section{Caching Model and Content Sharing Cost} 
\label{Caching_Model}
In this section, we discuss the caching policy and introduce our objective functions. 

\subsection{Caching Models}
We consider a probabilistic caching model for the content caching at the edge nodes, i.e. at both the D2D users and the BSs.
We define the probabilities that the BS $b_j$ ($j \in \mathcal{B}$) and the user $u_i$ ($i \in \mathcal{U}$) cache the content $f_k$ ($k \in \mathcal{F}$) as $\eta_{f_k}^{b_j}$ and $a_{f_k}^{u_i}$, respectively.  
The storage capacities of each BS and each user are denoted by $C_b$ and $C_d$, respectively. 
Hence, we have the constraints of $\sum_{k=1}^F \eta_{f_k}^{b_j} \leq C_b$ and $\sum_{k=1}^{F} a_{f_k}^{u_i} \leq C_d$, $\forall~ j$, $k$ and $i$.  
In the following, we consider the tagged user, which is defined as $u_i$; and the BS associated with the tagged user is the tagged BS (or serving BS) and is defined as $b_j$. 
Therefore, the remaining BSs are $b_{j'}$, where $j' \in \mathcal{B}\backslash \left\{j\right\}$.
Furthermore, the set of users in the coverage of $b_j$ is defined as $\mathcal{U}_{b_j} = \left\{1, \ldots, U_{b_j}\right\}$, where $U_{b_j}$ is the number of users including the tagged user.

\subsubsection{Heterogeneous caching model} 
In the heterogeneous caching placement case, heterogeneous user preferences as well as heterogeneous caching placement strategies are considered. 
In other words, the caching strategy at node $i$ is different from that of node $j$.
The probability of getting a content from self cache store $\mathrm{P}_o^{u_i}$, D2D neighbors $\mathrm{P}_d^{u_i}$, serving BS $\mathrm{P}_{b_j}^{u_i}$, neighbor BSs $\mathrm{P}_{B}^{u_i}$, locally $\mathrm{P}_l^{u_i}$ and from the cloud $\mathrm{P}_c^{u_i}$ are listed below in according order:
\begin{align}
    \mathrm{P}_o^{u_i} &= a_{f_k}^{u_i} \label{Pro_own}\\
    \mathrm{P}_d^{u_i} &= \left( 1 - a_{f_k}^{u_i} \right)  \left[1-\prod_{i' \in \mathcal{U}_{b_j} \backslash i}\left(1-a_{f_k}^{u_{i'}}\right)\right]\\
    \mathrm{P}_{b_j}^{u_i} & = \eta_{f_k}^{b_j} \prod_{i \in \mathcal{U}_{b_j}} \left(1 - a_{f_k}^{u_i} \right)\\
    \mathrm{P}_{\mathrm{B}}^{u_i}& = \left(1- \eta_{f_k}^{b_j} \right) \!\!\prod_{i \in \mathcal{U}_{b_j}}\!\!\!\!  \left(1 - a_{f_k}^{u_i} \right) \left[1 - \prod_{j' \in \mathcal{B} \backslash j}\left(1-\eta_{f_k}^{b_{j'}}\right) \right]\\
    \mathrm{P}_l^{u_i} &= 1 - \prod_{i\in \mathcal{U}_{b_j}} \left(1-a_{f_k}^{u_i}\right) \prod_{j \in \mathcal{B}}\left(1-\eta_{f_k}^{b_j}\right)\\
    \mathrm{P}_c^{u_i}  &= 1 - \mathrm{P}_l^{u_i}  = \prod_{i\in \mathcal{U}_{b_j}} \left(1-a_{f_k}^{u_i}\right) \prod_{j \in \mathcal{B}}\left(1-\eta_{f_k}^{b_j}\right). \label{Pro_Cloud}
\end{align}

\subsubsection{Homogeneous caching model}
In the homogeneous caching model, cache-enabled nodes store the same set of contents. 
Thus, the probabilities of storing a content into the cache-enabled nodes are equal for the same tier local nodes, i.e. $a_{f_k}^{u_1} =\dots = a_{f_k}^{U_{b_j}}$ and $\eta_{f_k}^{b_1} = \dots = \eta_{f_k}^{b_B}$, where $a_{f_k}^{u_i} \neq \eta_{f_k}^{b_j}$, $\forall i,j$. 
For simplicity, we get rid off the superscripts and denote the storing probabilities for D2D nodes and BSs as $a_{f_k}$ and $\eta_{f_k}$, respectively. 
Furthermore, we denote $U_c$ by the number of users in the cell.
We can rewrite (\ref{Pro_own}-\ref{Pro_Cloud}) as 
\begin{align}
	\mathrm{P}_o^{\mathrm{hom}} &= a_{f_k}.\\
	\mathrm{P}_d^{\mathrm{hom}} &= \left( 1 - a_{f_k} \right)  \left[1- \left(1-a_{f_k}\right)^{U_c-1} \right]. \\
	\mathrm{P}_{b_0}^{\mathrm{hom}} & = \left(1 - a_{f_k} \right)^{U_c}  \eta_{f_k}. \\
	\mathrm{P}_{\mathrm{B}}^{\mathrm{hom}} &= \left(1 - a_{f_k} \right)^{U_c} \left(1- \eta_{f_k} \right) \left[1 - \left( 1 - \eta_{f_k} \right)^{B-1} \right].\\ \!
	\mathrm{P}_l^{\mathrm{hom}} &= 1 - \!\left(1 - a_{f_k} \right)^{U_c} \left(1-\eta_{f_k} \right)^{B} .\\
	\mathrm{P}_c^{\mathrm{hom}} & = \left(1 - a_{f_k} \right)^{U_c} \left(1-\eta_{f_k} \right)^{B} . \!
\end{align}

\subsection{Content Sharing Cost}
\label{UPLEdge_Content_Sharing_Cost}
We now determine the cost of collaborating and sharing the contents among different nodes. 
We consider the two types of costs, namely (a) storage cost and (b) communication cost. 
For the communication cost, we consider transmission cost per bit per meter. 
If a content has a size of $S_f$ bits, then the transmission cost between D2D nodes residing $d$ meters apart is calculated as 
$\Lambda_d^{com} = S_f \times \delta_d \times d$,
where $\delta_d$ is the cost per byte transmission in case of D2D transmission. 
For simplicity, we consider equal storage cost - denoted by $\Lambda_{*}^{stor}$, for all nodes.
The cost of obtaining the content from node $*$  is
$\phi_{*} = \Lambda_{*}^{com} + \Lambda_{*}^{stor}$.
Here, $* \in \{\mathrm{C,BS,b_0,d}\}$, and $\phi_C, \phi_{BS},\phi_{b_0}$ and $\phi_d$ represent the costs of extracting a content from the cloud, the other BS in the same cluster, the serving BS and the other D2D nodes in the same cell, respectively. 
Furthermore, we assume that the transmission cost is zero, if the requested content is available in the storage of the requesting node itself. 
However, the storage cost must be included in this particular case.  
The relationship of the costs are presented in Proposition~\ref{Propostion_ContentShaingCost}.
\begin{prop} \label{Propostion_ContentShaingCost}
	In general, we assume that the costs of receiving the requested contents from various nodes satisfy the following constraint
	\begin{equation}
	\phi_{C} >> \phi_{BS}>> \phi_{b_0} >> \phi_{d}.
	\end{equation}
\end{prop}

Recall that the preference probability, $\rho_{f_k}^{u_i}$, is the long term probability that the user $u_i \in \{\mathcal{U}\}$ requests the content $f_k$. 
It is calculated in (\ref{Predicted_Preference}) based on the proposed LSTM model. 
The cost function for accessing a content $f_k$ by a tagged user is expressed as 
\begin{equation}
\label{cost}
\begin{aligned}
	\Xi_c &= \Lambda^{stor} \mathrm{P}_o^{u_i} + \phi_{d}  \mathrm{P}_d^{u_i} + \phi_{b}  \mathrm{P}_{b_0}^{u_i} + \phi_{BS} \mathrm{P}_{\mathrm{B}}^{u_i} + \phi_{C} \mathrm{P}_c^{u_i},
\end{aligned}
\end{equation}
where the first term is considered due to the fact that a requested content might need to be stored at the requesting user's own node. 
The second, third, fourth and fifth terms are considered for the cases of accessing the requested content from the neighboring D2D nodes, serving base station, other base stations of the cluster and cloud, respectively.

Next, the average cost for accessing the content among all the users and all the contents in a cluster is the weighted average of $\Xi_c$ in (\ref{cost}).
This quantity is calculated as 
\begin{equation}
\begin{aligned}
	\Xi_\pi &= \sum_{j=1}^{B} \sum_{i=1}^{U_{b_j}} \sum_{k=1}^F \rho_{f_k}^{u_i}  \frac{\Xi_c} {U}, \\			
\end{aligned}
\label{average_cost}
\end{equation}
where $j \in \{1, 2,\dots, B\}$ and $U$ represent the small cells and total number of users in a cluster, respectively, while
$U_{b_j}$ represents the number of users in the coverage of BS $b_j$.

\subsubsection{Heterogeneous caching model case} 

In case of heterogeneous caching placement, from (\ref{cost}) and (\ref{average_cost}), we rewrite $\Xi_\pi$ as 
\begin{equation}
\label{cost_function}
\begin{aligned}
	&\Xi_\pi^{\mathrm{het}} = \frac{1}{U}\sum_{j=1}^{B} \sum_{i = 1}^{U_{b_j}} \sum_{k=1}^F \rho_{f_k}^{u_i} \Bigg\{\Lambda^{stor} a_{f_k}^{u_i} + \phi_d \left(1-a_{f_k}^{u_i}\right) -\\
		& \quad A_1 \Bigg[\phi_d - \phi_b \eta_{f_k}^{b_j} - \phi_{BS} \left(1-\eta_{f_k}^{b_j}\right) + A_2 \left( \phi_{BS} - \phi_C \right)  \Bigg]  \Bigg\}, \\
\end{aligned}
\end{equation}
where $u_i$ and $b_j$ represent tagged user and serving base station, respectively. 
Recall that $\rho_{f_k}^{u_i}$ is calculated in (\ref{Predicted_Preference}).
$A_1$ and $A_2$ are calculated by $A_1 = \left(1 - a_{f_k}^{u_i} \right) \prod_{i' \in \mathcal{U}_{b_j} \backslash i } \left(1 - a_{f_k}^{u_{i'}} \right)$ and $A_2 = \left(1-\eta_{f_k}^{b_j}\right) \prod_{j' \in \mathcal{B} \backslash j } \left(1-\eta_{f_k}^{b_{j'}}\right)$.

We next formulate the optimization problem to minimize the content sharing cost, which is presented as
\begin{subequations}
	\begin{align}
	\mathbf{P_1:}\quad	& \underset{a_{f_k}^{u_i}, \eta_{f_k}^{b_j}}{\min} \quad  \Xi_\pi^{\mathrm{het}} \\
	& \quad \text{s. t.} \label{c1} \quad \sum_{k=1}^{F} a_{f_k}^{u_i} \leq C_d, \quad \forall i,k \\
	& \label{c2} \qquad \quad \sum_{k=1}^{F} \eta_{f_k}^{b_j} \leq C_b, \quad \forall k,j\\
	& \label{c3} \qquad \quad 0 \leq a_{f_k}^{u_i} \leq 1,~ 0 \leq \eta_{f_k}^{b_j} \leq 1, ~ \forall ~ i, ~ j ~\& k.
	\end{align}
	\label{opt1}
\end{subequations}
In problem $\mathbf{P_1}$, the constraints in (\ref{c1}) and (\ref{c2}) indicate that the total the contents cached at the node (i.e. a D2D node and a BS) must not excess the node's storage capacity. 
The constraint in (\ref{c3}) simply states that caching probabilities have to be in the range of $\left[0,1\right]$. 
Moreover, the cost function, $\Xi_\pi^{\mathrm{het}}$, is given in (\ref{cost_function}).

\subsubsection{Homogeneous caching model case} 

In case of homogeneous caching placement, from (\ref{cost}) and (\ref{average_cost}), we rewrite $\Xi_\pi$ as 
\begin{equation}
\label{cost_function_homogeneous}
\begin{aligned}
	&\!\!\!\!\Xi_\pi^{\mathrm{hom}} \!=\! \frac{1}{U}\sum_{j=1}^{B} \sum_{i = 1}^{U_c} \sum_{k=1}^F \rho_{f_k}^{u_i} \Bigg\{\Lambda^{stor} a_{f_k} + \phi_d \left(1-a_{f_k}\right) - \\ 
	& \qquad  B_1 \Bigg[\phi_d - \phi_b \eta_{f_k} - \phi_{BS} \left(1-\eta_{f_k}\right) + B_2 \left( \phi_{BS} - \phi_C \right)  \Bigg]  \Bigg\}, \\
\end{aligned}
\end{equation}
where $u_i$ represents the tagged user, while $U_c$ is the number of users in the cell.
Furthermore, $B_1$ and $B_2$ are $B_1 = \left(1 - a_{f_k} \right)^ {U_c}$ and $B_2 =  \left(1 - \eta_{f_k}\right)^{B}$. 
The detailed derivation of (\ref{cost_function_homogeneous}) is done by using some algebraic manipulations for (\ref{cost_function}). 

Here, we stress out the fact that all edge nodes (D2D nodes and BSs) are assumed to have an equal caching policy in homogeneous caching placement case \cite{zhang2015survey}.
Recall that $\rho_{f_k}^{u_i}$ is from (\ref{Predicted_Preference}).
Although we assume equal caching policy for all cache enabled nodes, we still consider heterogeneous content preferences of the users. 
Following the homogeneous notion, the optimization problem $\mathbf{P_1}$ is reformulated as 
\begin{subequations}
	\begin{align}
	\mathbf{P_2:}\qquad	& \underset{a_{f_k}, \eta_{f_k}}{\text{minimize}} \quad  \Xi_{\pi}^{\mathrm{hom}} \\
	& \text{s. t.} \label{cp1} \quad \sum_{k=1}^{F} a_{f_k} \leq C_d, \quad \forall i, k \\
	& \label{cp2}  \quad \quad \sum_{k=1}^F \eta_{f_k} \leq C_b, \quad \forall j, k\\
	& \label{cp3}  \quad \quad 0 \leq a_{f_k} \leq 1,~ 0 \leq \eta_{f_k} \leq 1, ~ \forall ~ i, ~ j ~\& k.
	\end{align}
	\label{opt2}
\end{subequations}
The constraints (\ref{cp1}) - (\ref{cp3}) are used for the same reasons as in problem $\mathbf{P_1}$.

\section{Joint Solver for the Objective Functions}
\label{Observations_Joint_Solver}

\subsection{Algorithm and Solver for the Joint Optimizations}

In this subsection, the proposed algorithms are presented to efficiently solve problems $\mathbf{P_1}$ and $\mathbf{P_2}$. 
Based on these above observation, the optimization problems $\mathbf{P_1}$ and $\mathbf{P_2}$ are not convex. 
Furthermore, user preferences vary dynamically over different time slots, which are captured by using the LSTM model. 
Considering these dynamics, this paper intends to capture the long term caching placement probabilities at the cache-enabled nodes. 
The significance of doing this is that a system administrator may need to know multiple time slots forecasts for the to-be-requested contents. 
If the binary cases\footnote{A binary case considers only $0$ or $1$. For example, if $a_{f_k}^{u_i}=0$, the content $f_k$ is not cached at the user node $u_i$.}, are considered, the obtained results are only for a single time slot. 
Instead, the goal of this paper is to optimize the caching placement probabilities for long-term cases. 
By doing that, two indicator functions, i.e. $\mathbb{I}_{f_k}^{u_i}(t_o)$ and $\mathbb{I}_{f_k}^{b_j}(t_o)$, are used to denote the cache placement indicators at users and BSs for time slot $t_0$, respectively. $\mathbb{I}_{f_k}^{u_i}(t_o) = 0$ and $\mathbb{I}_{f_k}^{u_i}(t_o) = 1$ indicate that content $f_k$ is not placed and placed into the cache storage of user $u_i$ for time slot $t_o$, respectively. 
This is essentially the binary case. 
This has to be considered for all optimization time slots and then finally, the cache placement probabilities are required to be calculated. 

Considering the above facts, the cache placement probabilities are calculated as follows:
\begin{equation}
\label{a_f_u_from_Indicator_func}
a_{f_k}^{u_i} = \frac{\sum_{t_o = N+1}^{N+N^{\mathrm{opt}}} \mathbb{I}_{f_k}^{u_i}(t_o) }{N^{\mathrm{opt}}}, ~ \forall ~i~ \text{and} ~k ,
\end{equation}
where $N^{\mathrm{opt}}$ is the total number of time slots for the optimization.
\begin{equation}
\label{eta_f_u_from_Indicator_func}
\eta_{f_k}^{b_j} = \frac{\sum_{t_0 = N+1}^{N+N^{\mathrm{opt}}} \mathbb{I}_{f_k}^{b_j}(t_o) }{N^{\mathrm{opt}}}, \forall ~j ~\text{and} ~k. 
\end{equation}
Now, to efficiently optimize the problems, necessary algorithms are proposed in what follows. 

\subsubsection{Algorithm and solver for heterogeneous caching placement}

In reality, solving the optimization problem in the case of heterogeneous caching placement strategy is more interesting and beneficial for a $\mathrm{CDN}$. However, the optimization problem $\mathbf{P_1}$ is very challenging and contains a large number of system parameters. Since the problem is not convex, it is extremely hard to get the optimal solutions. Therefore, heuristic algorithms are proposed to efficiently solve the joint optimization problem $\mathbf{P_1}$. Moreover, three scenarios are considered for placing the contents at the nodes for the heterogeneous case. The three sub-cases are - (a) \textit{collaborative greedy caching - base station first (non-overlapping)} (b) \textit{collaborative greedy caching - user first (non-overlapping)} and (c) collaborative greedy overlapping caching. 

\textbf{Collaborative greedy caching - base station first (non-overlapping)}: One way to think about this is to store as much content as possible. Therefore, the aim is to store the commonly preferred contents, $f_{\mathrm{com}}^C$ into the base stations cache storage first. No overlapping is considered in this case. In other words, uniquser content is stored at each cache-enabled nodes. First, the contents $f_{\mathrm{com}}^C$ are stored at the base stations. After that, if there is any place left, other preferred contents of the users (that are not already stored into the users' cache storage) of that respective cells are stored later on. Next, the users' preferred contents are placed into their cache storage. While doing so, it needs to be assured that there is no overlapping of similar content. After completing storing the contents at the user level, the base station's cache storage - given that there is actually some space left in its (BS) storage - is updated. 
The detailed procedures are listed in Alg. \ref{Greedy_Non_overlapping_BS_First_BS_Level}.

\begin{algorithm}
\small
	\caption{Collaborative greedy caching - base station first (non-overlapping)}
	\begin{algorithmic} [1]
		\For {each time slot, $t_o$ of the optimization of $\mathrm{P_1}$}
		\State \textbf{Input:} user content preference, $\hat{q}_{f_k,u_i}(t_o), ~\forall~u_i$
		\State calculate: $f_{\mathrm{com}}^{C} = f_{c_1}^{\mathrm{pref}} \cap f_{c_2}^{\mathrm{pref}} \cap f_{c_3}^{\mathrm{pref}} $ and $f_{\mathrm{com}}^{c_i c_j} = f_{c_i}^{\mathrm{pref}} \cap f_{c_j}^{\mathrm{pref}}$, $\forall~ U ~\& ~F$ 
		\State $f_{\mathrm{stored}} = []$ , $f_{\mathrm{com}}^{C_{\mathrm{rest}}} = []$, $C_{\mathrm{avail}}^b = []$, $f_{\mathrm{pref}}^{U_{\mathrm{rest}}}=[]$
		\For{each cell, $c_i \in \{C \}$ }
		\State store the common contents at first \Comment{$f_{\mathrm{com}}^{C}$ and $f_{\mathrm{com}}^{c_i c_j}$}
		\State update  $f_{\mathrm{stored}}$, $f_{\mathrm{com}}^{C_{\mathrm{rest}}}$ and $C_{\mathrm{avail}}^b$ \Comment{based on content popularity}
		\EndFor
		\State return $\mathbb{I}_{f_k}^{b_j}(t_o)$, $f_{\textrm{stored}}$ and $C_{\mathrm{avail}}^b$ 
		\For {each cell, $c_i \in \{C \}$}
		\For {$\forall$ $u_i \in \{U_c\}$}
		\For {$\forall$ $f_k \in \{f_{u_i}^{\mathrm{pref}}\}$}
		\If {$f_{u_i}^{\mathrm{pref}} \notin f_{\mathrm{stored}}$ \&\& $C_d \neq $ full}
		\State $\mathbb{I}_{f_k}^{u_i}(t_o) \leftarrow 1 $
		\State update $f_{\mathrm{stored}}$, $f_{\mathrm{pref}}^{U_{\mathrm{rest}}}$ 
		\EndIf 
		\EndFor
		\EndFor 
		\If {$C_{\mathrm{avail}}^b$ $\neq 0$}
		\State fill out the storage with $f_{\mathrm{pref}}^{U_{\mathrm{rest}}}$
		\EndIf
		\EndFor
		\State \textbf{return}   $\mathbb{I}_{f_k}^{u_i}(t_o)$ and $\mathbb{I}_{f_k}^{b_j}(t_o)$
		\EndFor
		\State \textbf{calculate} $\bar{a}_{f_k}^{u_i}$ and $\bar{\eta}_{f_k}^{b_j}$, $\forall~u_i~ \& ~b_j$ using equations (\ref{a_f_u_from_Indicator_func}-\ref{eta_f_u_from_Indicator_func})
		\State Return $\Xi_{\pi}^{\mathrm{het}}$    
	\end{algorithmic} \label{Greedy_Non_overlapping_BS_First_BS_Level}
\end{algorithm}

\textbf{Collaborative greedy caching - user first (non-overlapping)}: In this case, a non-overlapping cache placement strategy is considered. Here, user cache storage is filled with the most requested and popular content first. Then, the residual contents are placed at the base stations. It is worth mentioning that it is very similar to the \textit{collaborative greedy caching - base station first (non-overlapping)} case. However, the difference is - the contents are placed at the user level first. For brevity, we do not present the repetitive algorithm here.

\textbf{Collaborative greedy overlapping caching}: In this case, a completely greedy caching mechanism is adopted. As the cost of getting the requested content from other nodes is higher than storing the content at the requester node, the aim of this algorithm is to place as many to-be-requested content as possible into the requester cache storage. Recall that the prediction model can predict what content a user will request ahead of time. Therefore, it makes sense to polish the caching policy based on the user's preferences. Using the forecast information, the to-be-requested content by the users is placed into their cache storage for each time slot. This gives the indicator functions $\mathbb{I}_{f_k}^{u_i}(t_o)$s. Finding the indicator functions then gives the long term cache placement probabilities. For the base station's cache storage, the remaining contents are placed based on their popularity profile. Finally, the caching placement probabilities $a_{f_k}^{u_i}$ and $\eta_{f_k}^{b_j}$ are calculated using equations (\ref{a_f_u_from_Indicator_func}) and (\ref{eta_f_u_from_Indicator_func}), respectively. The detailed algorithm for this case is presented in Alg. \ref{Collaborative_greedy_overlapping_caching}.

\begin{algorithm}[t]
\small
\caption{Collaborative Greedy overlapping Caching}
\begin{algorithmic}[1]
	\For {each time slot, $t_o$ of the optimization of $\mathrm{P_1}$}
		\State \textbf{input:} predicted user content preference, $\hat{q}_{f_k,u_i}(t_o)$
		\For{each cell, $j \in \mathcal{B}$}
			\State $f_{\mathrm{stored}}^u=[]$, $f_u^{\mathrm{rest}}=[]$, ${C}_d^{\mathrm{avail}} = []$, $\mathit{Checksum} = 0$
			\For {each user, $u_i \in \{U_{b_j}\}$}
				\State find $f_\mathrm{pref}$ and sort $f_\mathrm{pref}$ based on $\hat{q}_{f_k,u_i}(t_o)$
				\If {$len(f_{\mathrm{pref}}) > {C}_d$ }
					\State $\mathbb{I}_{f_k}^{u_i}(t_o) \leftarrow index(f_{\mathrm{pref}}[0:C_d])$
					\State $f_{\mathrm{stored}}^u.\mathrm{append}(index(f_{\mathrm{pref}}[0:C_d]))$
					\State $f_u^{\mathrm{rest}} \leftarrow index(f_{\mathrm{pref}}[C_d:end])$
				\Else  \Comment{$len(f_{\mathrm{pref}}) \leq C_d$}
					\State  $\mathbb{I}_{f_k}^{u_i}(t_o) \leftarrow index(f_{\mathrm{pref}})$
					\State $f_{\mathrm{stored}}^u.\mathrm{append}(index(f_{\mathrm{pref}}))$
					\State $S_{\mathrm{avail}} = C_d - len(f_{\mathrm{pref}}) $
					\State ${C}_d^{\mathrm{avail}}. \mathrm{append}(S_{\mathrm{avail}})$
				\EndIf
			\EndFor
			\State find the index of $f_u^{\mathrm{rest}}$ and $\hat{q}_{f_k,u_i}(t_o)$
			\State $f_u^{\mathrm{rest_{up}}} \leftarrow sort(f_u^{\mathrm{rest}})$ \Comment{descending order}
			\If {$len(f_u^{\mathrm{rest_{up}}}) > \sum_{i=1}^{U_{b_j}}({C}_d^{\mathrm{avail}})$}
				\For {$\forall$ $u_i$ in which ${C}_d^{avail} \neq 0$} 				\label{label1}
					\State $\mathbb{I}_{f_k}^{u_i}(t_o).\mathrm{extend} (f_u^{\mathrm{rest_{up}}}[0:{C}_d^{\mathrm{avail}}])$, $\forall ~$item$ ~$in$ ~f_u^{\mathrm{rest_{up}}} \notin f_{\mathrm{stored}}^u$ \Comment{if in $f_{\mathrm{stored}}^u$, store the next popular one and delete it from $f_u^{\mathrm{rest_{up}}}$}
					\State $f_u^{\mathrm{rest_{up}}} = f_u^{\mathrm{rest_{up}}}[{C}_d^{\mathrm{avail}}: end]$
				\EndFor  \label{label2}
				\State set $\mathit{Checksum} += 1$
			\Else 
				\State repeat steps (\ref{label1}-\ref{label2}), if any storage is yet left consider storing the most popular content in that cell
			\EndIf 
			\If {$\mathit{Checksum} \neq 0$}
				\If {$len(f_u^{\mathrm{rest}_{\mathrm{up}}}) > C_b$ }
					\State $\mathbb{I}_{f_k}^{b_j}(t_o) \leftarrow 1$, $\forall$ $f_k \in f_u^{\mathrm{rest}_{\mathrm{up}}}[0:C_b]$
				\Else 
					\State $\mathbb{I}_{f_k}^{b_j}(t_o) \leftarrow 1$, $\forall$ $f_k \in f_u^{\mathrm{rest}_{\mathrm{up}}}$ \label{Label3}
					\State fill out the BS storage (if any space left after step \ref{Label3}) with the most popular content of the cell \label{Label4}
				\EndIf
			\Else 
				\State repeat step (\ref{Label4})
			\EndIf 
		\EndFor 
	\EndFor
	\State \textbf{calculate} $\bar{a}_{f_k}^{u_i}$ and $\bar{\eta}_{f_k}^{b_j}$, $\forall~u_i~ \& ~b_j$ using equations (\ref{a_f_u_from_Indicator_func}-\ref{eta_f_u_from_Indicator_func})
	\State Return $\Xi_{\pi}^{\mathrm{het}}$   
\end{algorithmic} \label{Collaborative_greedy_overlapping_caching}
\end{algorithm}

\subsubsection{Algorithm and solver for homogeneous caching placement}

To tackle the complexity, one may consider homogeneous caching placement. 
Recall that in homogeneous caching policy, all nodes in the similar tier places same content into their caches. 
As the problem $\mathbf{P_2}$ is not a convex problem, it is hard to get the optimal solution. 
We again propose a heuristic algorithm to efficiently solve the joint optimization problem. 
In particular, we consider that all D2D nodes in the same cell follow homogeneity, while placing the content. 
Similarly, all BSs in the same cluster follow homogeneity.
However, the preferences of the users are not homogeneous. 
Each user has different preference than others. 
Therefore, it is expected that the system performance will degrade, while the complexity will be definitely reduced. 
It implies that there is a trade off between performance and complexity.

In the homogeneous caching model, for all optimization time slots, contents are placed at the user level first. Then, the residual contents are stacked and sorted (based on their popularity). Finally, the base stations' cache stores are filled out with the most popular content. 
The detailed procedure is presented in Alg. \ref{Collaborative_Caching_Algorithm_Homogeneous_Case}. 

\begin{algorithm}
\small
	\caption{Collaborative Edge Caching Algorithm: Homogeneous Case}
	\begin{algorithmic}[1]
		\For {each time slots, $t_o$ of the optimization of $\mathrm{P_2}$}
		\For {each cell, $c \in \{C\}$}
		\State calculate $\Omega_{f_k}^c = \sum_{i=1}^{U_c} \hat{q}_{f_k,u_i}(t_o)$
		\State find and sort $f_{c}^{\mathrm{pref}}$ using $\Omega_{f_k}^c$ \Comment{descending order}
		\For {$\forall~u_i \in \{U_c\}$}
		\State $\mathbb{I}_{f_k}^{u_i}(t_o) \leftarrow 1$, $\forall~f_k \in ~f_{c}^{\mathrm{pref}}[0:\mathcal{C}_d]$
		\State $f_c^{\mathrm{stored}}.\mathrm{append}[index(f_{c}^{\mathrm{pref}}[0:\mathcal{C}_d])]$
		\State $f_c^{\mathrm{residual}}.\mathrm{append}[index(f_{c}^{\mathrm{pref}}[\mathcal{C}_d:end])]$
		\EndFor
		\State $f_{\mathrm{Cell}}^{\mathrm{residual}}.\mathrm{append}(f_c^{\mathrm{residual}}) $
		\EndFor
		\State calculate $\Omega_{f_k}^{\mathrm{Cell}} = \sum_{c=1}^{C}\sum_{i=1}^{U_c} \hat{q}_{f_k,u_i}(t_o) $
		\State sort $f_{\mathrm{Cell}}^{\mathrm{residual}}$ based on $\Omega_f^{\mathrm{Cell}}$ \Comment{descending order}
		\For {$\forall~b_j\in \{B\}$}
		\State $\mathbb{I}_{f_k}^{b_j}(t_o) \leftarrow 1$, $\forall~f_k\in ~f_{\mathrm{Cell}}^{\mathrm{residual}}[0:C_b]$
		\State $f_b^{\mathrm{stored}}.\mathrm{append}[index(f_{\mathrm{Cell}}^{\mathrm{residual}}[0:C_b])]$
		\EndFor
		\EndFor 
		\State calculate $\bar{a}_{f_k}$ and $\bar{\eta}_{f_k}$ using equations (\ref{a_f_u_from_Indicator_func}-\ref{eta_f_u_from_Indicator_func})
		\State calculate and \textbf{return}: $\Xi_\pi^{\mathrm{hom}}$
	\end{algorithmic} \label{Collaborative_Caching_Algorithm_Homogeneous_Case}
\end{algorithm}

\section{Results and Discussion}
\label{Result}

The simulation parameter setting is given as follows:
total number of content, $F=225$ , total number of users, $U=45$; total number of BSs in a cluster, $B=3$; total number of users under a serving BS is $15$; $C_b$ is in the range of $\left[5,14\right]$; $C_d$ is in the range of $\left[1,4\right]$; number of historical time slots, $t \in \{1,2, \dots, N\}$, $N = 250$; number of optimization time slots, $t_0 = \{N+1, N+2, \dots, N+N^{\mathrm{opt}}\}$, $N^{\mathrm{opt}} = 50$; $\Lambda^{stor}$ = 2000; $\{\Lambda_{d}^{com}, \Lambda_{b_0}^{com}, \Lambda_{BS}^{com}, \Lambda_{C}^{com} \} =\{100, 500, 1000, 5000\}$. 
The following simulation results validate the theoretical findings.

After generating the initial content request number following Alg.~ \ref{Synthetic_Data_Set_Generation}, correlated request numbers are generated using $n_{u_i f_k} (t) =  n_{u_i f_k} (t_{\mathrm{int}}) + \sum_{n=1}^{\infty} A_n \sin(nt) + \epsilon(t)$, where $n_{u_i f_k} (t_{\mathrm{int}})$ represents initial generated number for time slot $1$, $t$ represents rest of the time slots for which the correlated data are being generated, $A$ represents amplitude and $\epsilon (t)$ is $\mathrm{Normal}$ random variable with $\mathrm{mean}$ 0 and $\mathrm{variance}$ 1. We consider $A_n =1$, $n=1,2,3$ and $t=2,3,\dots,250$ for our simulation. 
Also, as the requested incident number is non-negative and integer valued, necessary replacement of any negative number with $0$ and rounding are performed. 
We stress out that the proposed LSTM is a powerful solution and can be readily extended for any other kind of co-related data generation process. 
Given enough data samples, our proposed method is capable of predicting dynamic user preferences efficiently.

\begin{figure*}
	\centering
	\subfloat[] []{\label{Prediction_for_User_01_Content_112} \includegraphics[trim=10 170 10 180, clip, width= 0.5 \textwidth, height = 0.25 \textheight]{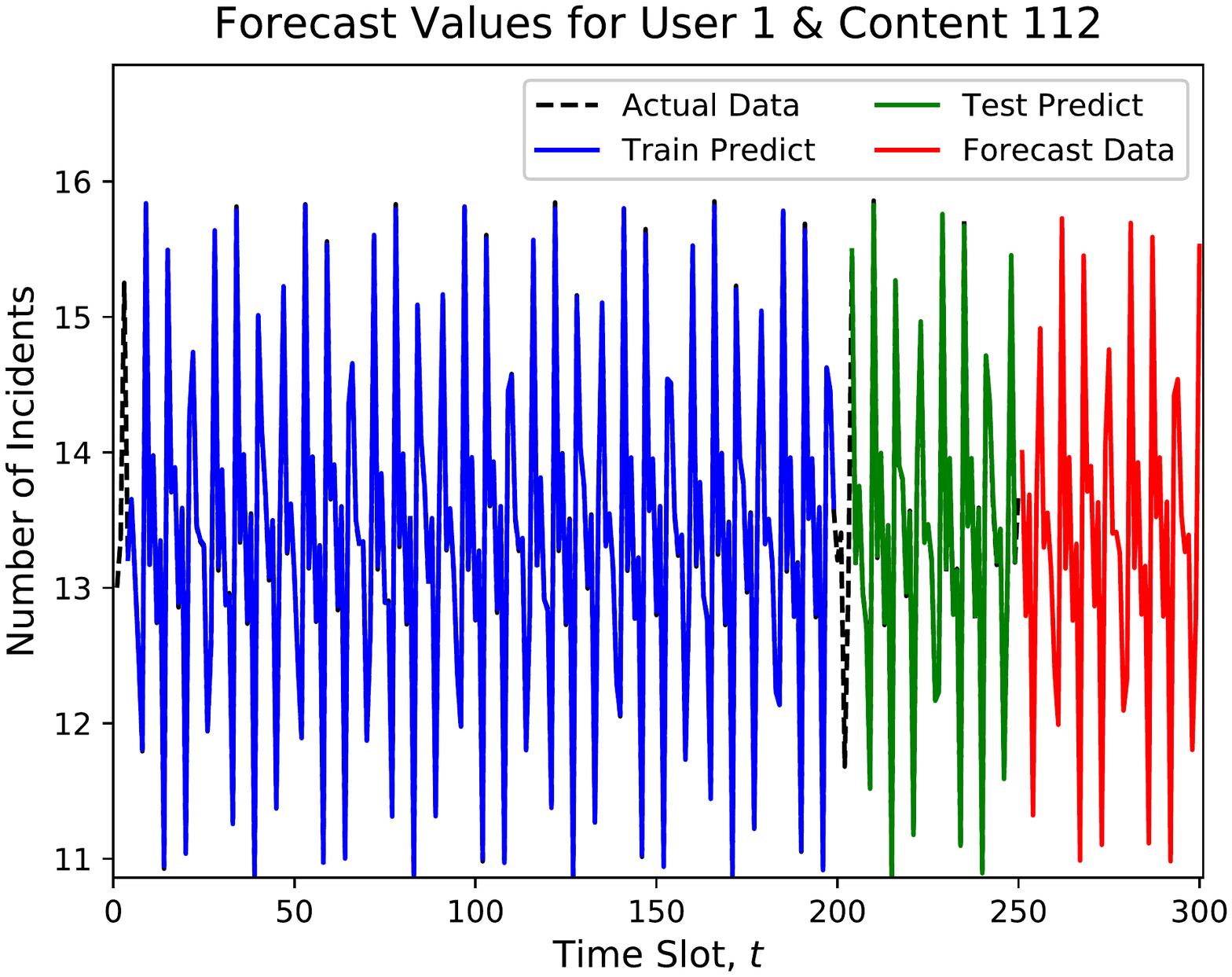}}
	\subfloat[] [] {\label{Forecasted_Activity_Levels} \includegraphics[trim=10 170 10 180, clip, width = 0.5 \textwidth, height = 0.25 \textheight]{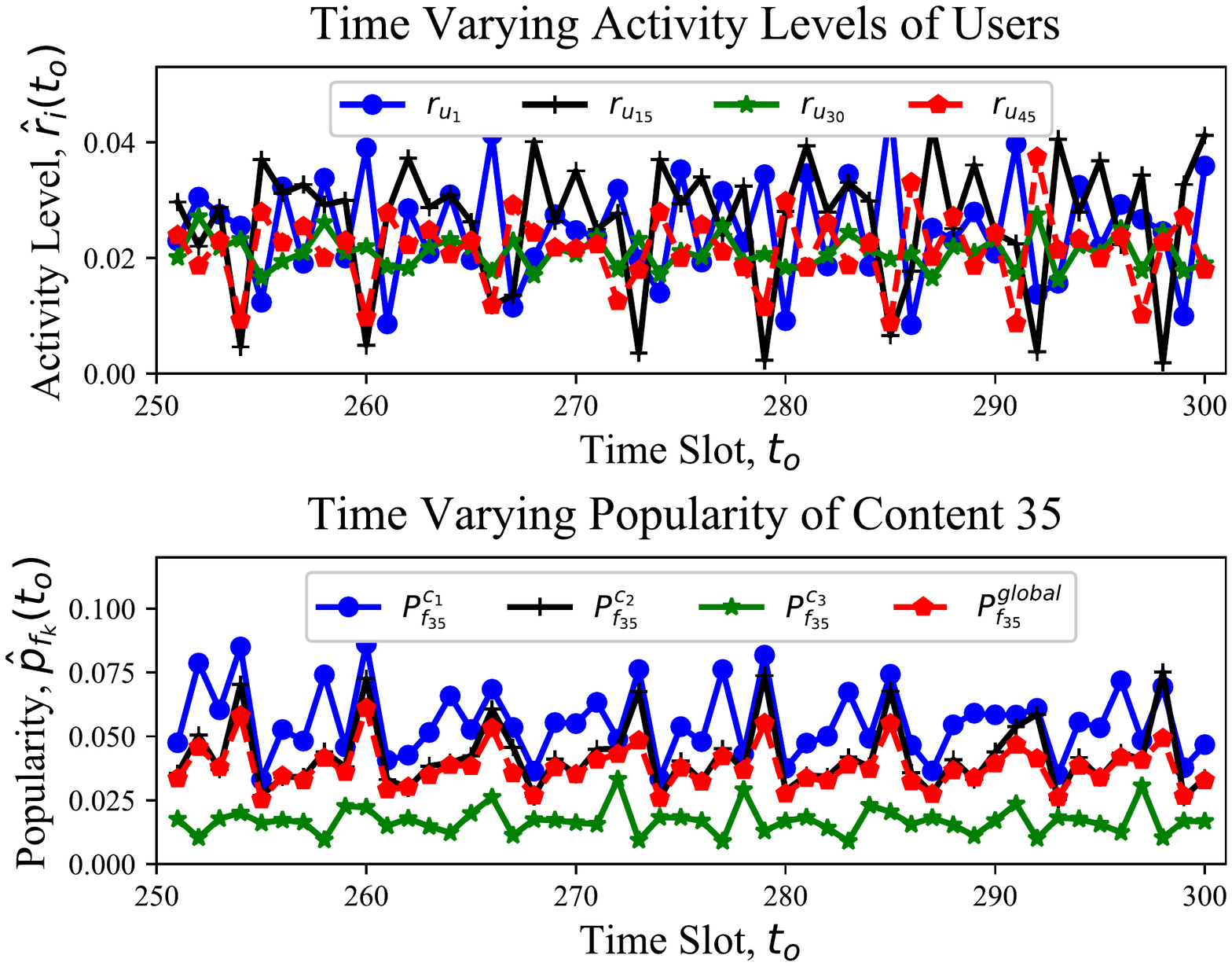}} \\
	\subfloat[] [] {\label{Comparision_CostFunctions_Stat_Dynamic} \includegraphics[trim=10 170 10 180, clip, width =  \textwidth, height = 0.5 \textheight]{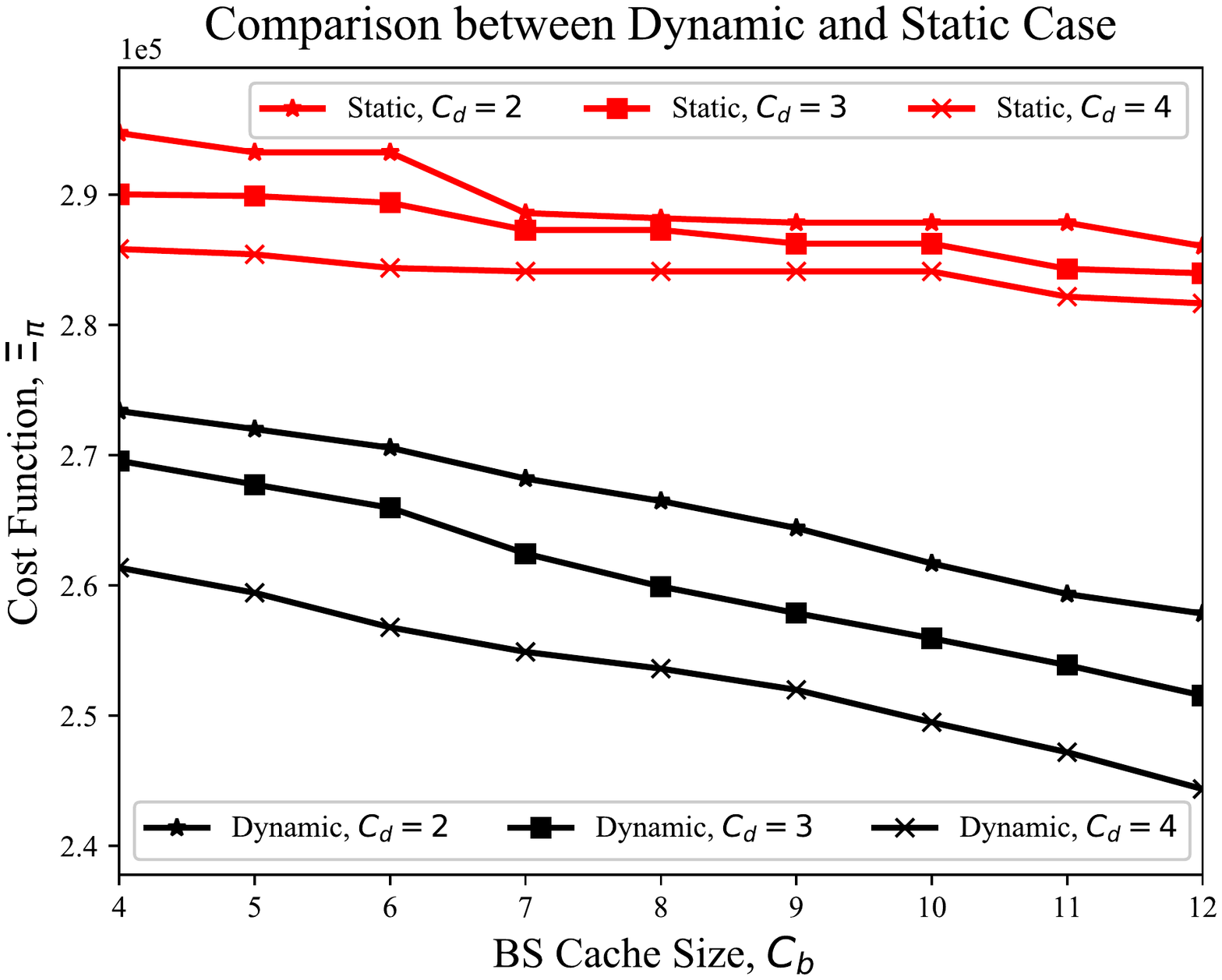}}
	\caption{(a) Predicted value for user $1$ and content $112$; (b) Time varying nature of users' content preferences and activity levels; and (c) comparison between existing static \cite{8410473} and proposed dynamic case} \vspace{-0.15 in}
\end{figure*}

Now, using the proposed prediction model in Alg. \ref{Predicting_values_Using_LSTM}, the contents that will be requested in the next time slot by the users are sequentially predicted. 
The prediction made by this model for the most popular content of user $1$ is shown in Fig.~\ref{Prediction_for_User_01_Content_112}. 
To show the temporal dynamics over time slots, we present our results for some selected users from all cells. 
Here, we capture the dynamics for all users and all contents in all time slots. 
We illustrate only a sample of how popularity of the contents and activity of the users change over time in Fig.~\ref{Forecasted_Activity_Levels}. 
Using these values, we measure the content preference probabilities ($\rho_{f_k}^{u_i}$) of the users. 
We then use these results for the caching policy designing in the next sub-section.

\begin{figure*}
	\centering
	\subfloat[] [\label{Cost4differentBSCacheSizeWithCd4}] {\includegraphics[trim=10 170 10 180, clip, width = 0.5 \textwidth] {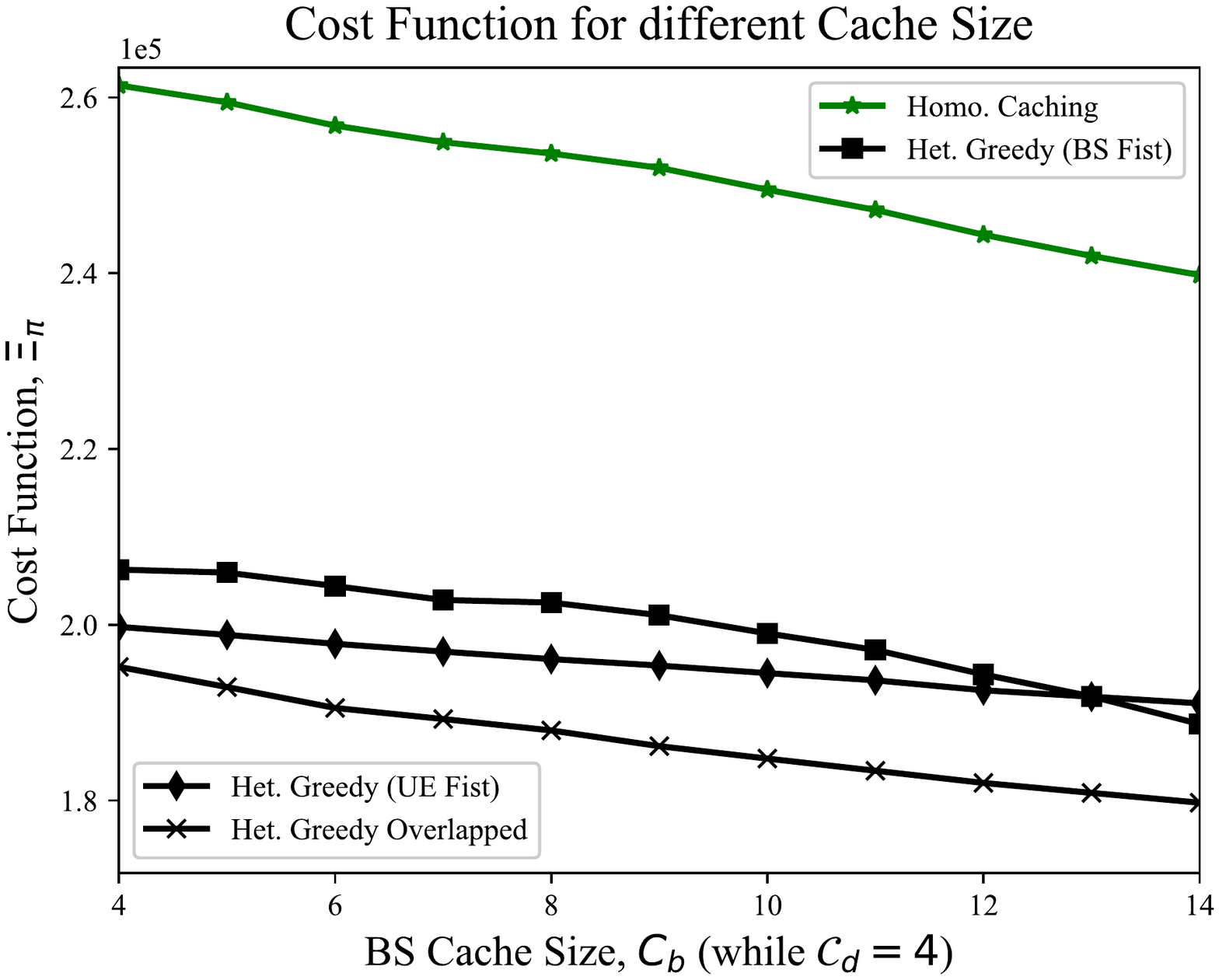}}
	\subfloat[] [\label{Cost4differentuserCacheSizeWithCb12}] {\includegraphics[trim=10 170 10 180, clip, width = 0.5 \textwidth] {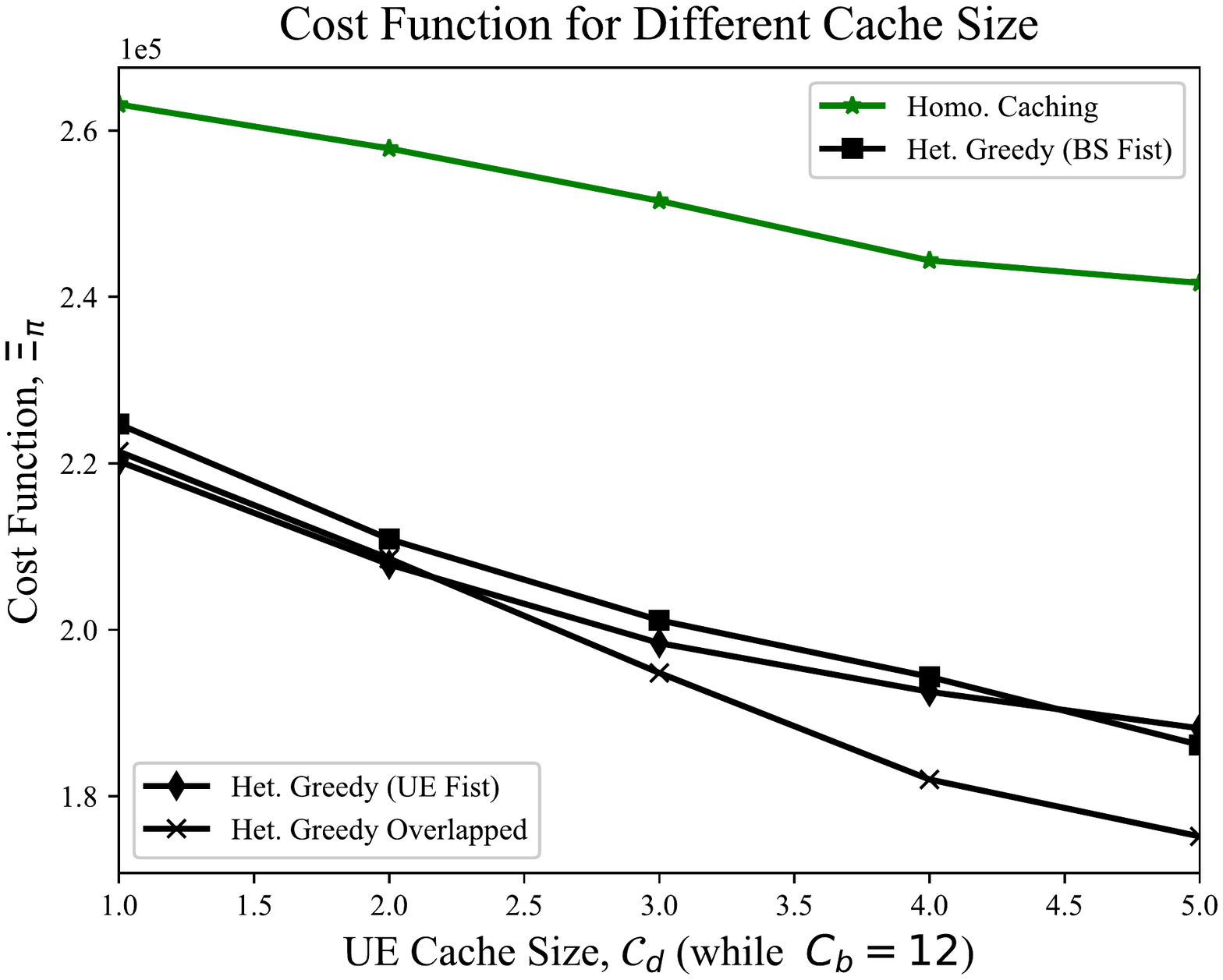}} 
	\caption{ (a) Cost functions for different BS cache sizes; and (b) Cost functions for different user cache sizes} \vspace{-0.15 in}
\end{figure*}

We firstly compare the performance between the static caching placement \cite{8410473} and the proposed dynamic prediction-based caching strategy in Fig.~\ref{Comparision_CostFunctions_Stat_Dynamic}. 
Particularly, we take the static caching model of \cite{8410473} with the homogeneous caching case and compare the results with our proposed scheme.
In the static case \cite{8410473}, there is no information about the temporal dynamics of the user preferences and activity levels for all time slots $t_o$. 
Therefore, the caching placement probabilities are the same in all time slots. 
However, all of the temporal dynamics are well captured in our proposed scheme. 
Hence, the system administrator knows precisely at what time, what contents might be requested by the users. 
Therefore, the optimal caching placement can be performed based on the requirement.  
In Fig \ref{Comparision_CostFunctions_Stat_Dynamic}, it is quite apparent that the proposed dynamic prediction-based caching strategy outperforms static caching placement \cite{8410473}. 
Therefore, in the following, we only show comparisons among our proposed caching schemes.



We firstly compare the performance between the static caching placement and the proposed dynamic prediction-based caching strategy in Fig.~\ref{Comparision_CostFunctions_Stat_Dynamic}. 
In the static case, there is no information about the temporal dynamics of the user preferences and activity levels for all time slots $t_o$. 
Therefore, the caching placement probabilities are the same in all time slots. 
However, all of the temporal dynamics are well captured in our proposed scheme. 
Hence, the system administrator knows precisely at what time, what contents might be requested by the users. 
Furthermore, the load coming from all of the users are also known to the system administrator. 
Therefore, the optimal caching placement can be performed based on the requirement.  
In Fig \ref{Comparision_CostFunctions_Stat_Dynamic}, it is quite apparent that the proposed dynamic prediction-based caching strategy outperforms static caching placement. 
Therefore, the proposed LSTM model, and its obtained results will be used for conducting various performance analysis in the following.

In Fig. \ref{Cost4differentBSCacheSizeWithCd4}, we illustrate the cost performance of our proposed schemes, where $C_d=4$ and $C_b$ varies in the range of $\left[4,14\right]$. 
We can easily see that the proposed \textit{greedy overlapping caching} placement performs significantly better than all the other cases. 
However, a critical observation is when $C_b$ increases, the performance of the collaborative \textit{greedy caching - BS first (non-overlapping)} is better than that of collaborative \textit{greedy caching - user first (non-overlapping)}. 
For example, when $C_b=13$, the performance of these two algorithms are approximately identical; when $C_b =14$, the performance of the collaborative \textit{greedy caching - base station first (non-overlapping)} case is visibly better than the later one. This is duser to the fact that when $C_b$ increases, more contents can be stored at the BSs first. 
Then, the rest of the popular contents are stored at the users. 
While performing the cache placement, the user's preferred leftover contents are being stored at the respective user node first. 
Therefore, more contents are being stored at the users. 
Whereas in collaborative \textit{greedy caching - user first (non-overlapping)}, the common contents are stored at the users first and then the BS cache storage is filled out. 
As the user cache size is fixed, the user may not store its own preferred contents as the common contents that are requested by all users in the respective cell. 
However, the proposed collaborative \textit{greedy overlapping caching} algorithm outperforms all the others in this case.
Since the user cache storage size is at a moderate level and the preferred contents are stored at its self cache storage first.
Furthermore, if the cache storage of the users is significantly small, the performances of the proposed \textit{greedy overlapping caching} and \textit{greedy caching - user first (non-overlapping)}, are similar. 
Because the most popular and common contents for the users are stored into either the requester self cache storage or nearby neighbor nodes first. 
It means that the cost is either only due to the storage cost (if stored in the own cache store) or a small transmission cost for obtaining from the neighbors.

Finally, Fig.~\ref{Cost4differentuserCacheSizeWithCb12} illustrates the cost performance vs the user cache size for $C_b = 12$. 
When $C_d$ is small, the performances of the \textit{greedy - (a) user first and (b) overlapping caching} are nearly similar. 
Because the user's most preferred contents are firstly stored into its cache. 
However, as the cache size of the user increases, as expected, the proposed \textit{greedy overlapping caching} policy outperforms all other caching placement strategies. 
Note that we have the similar critical observation, as described in Figs.~ \ref{Cost4differentBSCacheSizeWithCd4}, \ref{Cost4differentuserCacheSizeWithCb12}. 
That is the point of $C_d= 4.5$ for the collaborative \textit{greedy - (1) BS first (non-overlapping) and (2) user first (non-overlapping)}. 
Whereas, as deemed, the proposed \textit{greedy overlapping caching} algorithm outperforms the others.

\section{Conclusion}
\label{Conclu}
In a content delivery network, obtaining accurate content popularity prediction is immensely influential yet a difficult task. 
Using the LSTM model, we have successfully captured the temporal dynamics of the user preferences and their activity levels. 
With the theoretical analysis and experimental simulation, we demonstrated that the system performance highly depends on the prediction of the content dynamics and popularity.  
We furthermore made fair comparisons among different cache placement strategies and concluded that the proposed greedy overlapping caching mechanism outperforms other alike caching schemes.

\bibliography{Reference}

\begin{thebibliography}{10}
\providecommand{\url}[1]{#1}
\csname url@samestyle\endcsname
\providecommand{\newblock}{\relax}
\providecommand{\bibinfo}[2]{#2}
\providecommand{\BIBentrySTDinterwordspacing}{\spaceskip=0pt\relax}
\providecommand{\BIBentryALTinterwordstretchfactor}{4}
\providecommand{\BIBentryALTinterwordspacing}{\spaceskip=\fontdimen2\font plus
\BIBentryALTinterwordstretchfactor\fontdimen3\font minus
  \fontdimen4\font\relax}
\providecommand{\BIBforeignlanguage}[2]{{%
\expandafter\ifx\csname l@#1\endcsname\relax
\typeout{** WARNING: IEEEtran.bst: No hyphenation pattern has been}%
\typeout{** loaded for the language `#1'. Using the pattern for}%
\typeout{** the default language instead.}%
\else
\language=\csname l@#1\endcsname
\fi
#2}}
\providecommand{\BIBdecl}{\relax}
\BIBdecl

\bibitem{Pervej_Cost}
{M. F. {Pervej}}, {L. T. {Tan}}, and {R. Q. {Hu}}, ``User preference learning
  aided collaborative edge caching for small cell networks,'' in \emph{{Proc.
  IEEE Globecom}}, Dec. 2020.

\bibitem{5978416}
A.~Fehske, G.~Fettweis, J.~Malmodin, and G.~Biczok, ``The global footprint of
  mobile communications: The ecological and economic perspective,'' \emph{IEEE
  Commun. Mag.}, vol.~49, no.~8, pp. 55--62, August 2011.

\bibitem{bandyopadhyay2011internet}
D.~Bandyopadhyay and J.~Sen, ``Internet of things: Applications and challenges
  in technology and standardization,'' \emph{Wireless Personal Commun.},
  vol.~58, no.~1, pp. 49--69, 2011.

\bibitem{nayak20206g}
S.~Nayak and R.~Patgiri, ``6g: Envisioning the key issues and challenges,''
  \emph{arXiv preprint arXiv:2004.04024}, 2020.

\bibitem{8447267}
L.~T. Tan and R.~Q. Hu, ``Mobility-aware edge caching and computing in vehicle
  networks: A deep reinforcement learning,'' \emph{IEEE Trans. Veh. Tech.},
  vol.~67, no.~11, pp. 10\,190--10\,203, Nov 2018.

\bibitem{8618350}
L.~T. Tan, R.~Q. Hu, and L.~Hanzo, ``Twin-timescale artificial intelligence
  aided mobility-aware edge caching and computing in vehicular networks,''
  \emph{IEEE Trans. Veh. Tech.}, pp. 1--1, 2019.

\bibitem{8410473}
L.~T. Tan, R.~Q. Hu, and Y.~Qian, ``D2d communications in heterogeneous
  networks with full-duplex relays and edge caching,'' \emph{IEEE Trans. Ind.
  Informat.}, vol.~14, no.~10, pp. 4557--4567, Oct 2018.

\bibitem{8169053}
S.~Zhang, P.~He, K.~Suto, P.~Yang, L.~Zhao, and X.~Shen, ``Cooperative edge
  caching in user-centric clustered mobile networks,'' \emph{IEEE Trans. Mobile
  Comput.}, vol.~17, no.~8, pp. 1791--1805, Aug 2018.

\bibitem{ye2018smart}
F.~Ye, Y.~Qian, and R.~Q. Hu, \emph{Smart Grid Communication Infrastructures:
  Big Data, Cloud Computing, and Security}.\hskip 1em plus 0.5em minus
  0.4em\relax John Wiley \& Sons, 2018.

\bibitem{pervej2020eco}
{M. F. {Pervej}} and {S.-C. {Lin}}, ``{Eco-Vehicular} edge networks for
  connected transportation: A distributed multi-agent reinforcement learning
  approach,'' in \emph{Proc. IEEE VTC2020-Fall}, Oct. 2020.

\bibitem{pervej2020dynamic}
M.~F. Pervej and S.-C. Lin, ``Dynamic power allocation and virtual cell
  formation for {Throughput-Optimal} vehicular edge networks in highway
  transportation,'' in \emph{Proc. IEEE ICC Workshops}, June 2020.

\bibitem{6787081}
N.~Golrezaei, P.~Mansourifard, A.~F. Molisch, and A.~G. Dimakis, ``Base-station
  assisted device-to-device communications for high-throughput wireless video
  networks,'' \emph{IEEE Trans. Wireless Commun.}, vol.~13, no.~7, pp.
  3665--3676, July 2014.

\bibitem{6600983}
K.~Shanmugam, N.~Golrezaei, A.~G. Dimakis, A.~F. Molisch, and G.~Caire,
  ``Femtocaching: Wireless content delivery through distributed caching
  helpers,'' \emph{IEEE Trans. on Inf. Theory}, vol.~59, no.~12, pp.
  8402--8413, Dec 2013.

\bibitem{7944647}
J.~Song, M.~Sheng, T.~Q.~S. Quek, C.~Xu, and X.~Wang, ``Learning-based content
  caching and sharing for wireless networks,'' \emph{IEEE Trans. on Commun.},
  vol.~65, no.~10, pp. 4309--4324, Oct 2017.

\bibitem{8977517}
L.~{Jiang} and X.~{Zhang}, ``Cache replacement strategy with limited service
  capacity in heterogeneous networks,'' \emph{IEEE Access}, vol.~8, pp.
  25\,509--25\,520, 2020.

\bibitem{8647483}
J.~{Chen}, W.~{Xu}, N.~{Cheng}, H.~{Wu}, S.~{Zhang}, and X.~{Shen},
  ``Reinforcement learning policy for adaptive edge caching in heterogeneous
  vehicular network,'' in \emph{Proc. GLOBECOM}, Dec. 2018.

\bibitem{7775114}
S.~{Müller}, O.~{Atan}, M.~{van der Schaar}, and A.~{Klein}, ``Context-aware
  proactive content caching with service differentiation in wireless
  networks,'' \emph{IEEE Trans. Wireless Commun.}, vol.~16, no.~2, pp.
  1024--1036, Feb 2017.

\bibitem{7524381}
S.~{Li}, J.~{Xu}, M.~{van der Schaar}, and W.~{Li}, ``Popularity-driven content
  caching,'' in \emph{In Proc. IEEE INFOCOM 2016}, 2016.

\bibitem{cha2007tube}
M.~Cha, H.~Kwak, P.~Rodriguez, Y.-Y. Ahn, and S.~Moon, ``I tube, you tube,
  everybody tubes: analyzing the world's largest user generated content video
  system,'' in \emph{Proc. 7th ACM SIGCOMM}, 2007.

\bibitem{8425746}
B.~{Chen} and C.~{Yang}, ``Caching policy for cache-enabled d2d communications
  by learning user preference,'' \emph{IEEE Trans. Commun.}, vol.~66, no.~12,
  pp. 6586--6601, Dec 2018.

\bibitem{zhang2015survey}
M.~Zhang, H.~Luo, and H.~Zhang, ``A survey of caching mechanisms in
  information-centric networking,'' \emph{IEEE Commun. Surveys \& Tuts.},
  vol.~17, no.~3, pp. 1473--1499, 2015.

\end{thebibliography}
\bibliographystyle{IEEEtran}

\end{document}